\documentclass[10pt,twocolumn,letterpaper]{IEEEtran}
\usepackage{amsmath,epsfig}
\usepackage{algorithm}
\usepackage{bm}
\usepackage{tzplot}
\usepackage[usenames,dvipsnames]{pstricks}
\usepackage{epsfig}
\usepackage{pst-grad} 
\usepackage{pst-plot} 
\usepackage{subfigure}
\usepackage[capitalize]{cleveref}

\usepackage{subfigure}
\usepackage[T1]{fontenc}
\usepackage{amsmath}
\usepackage{graphics}
\include{psfig}
\usepackage{epsfig}
\usepackage{array}
\usepackage{psfrag}
\usepackage{amssymb}
\usepackage{color}
\usepackage{pstricks,pst-node,pst-text,pst-3d,pst-plot}
\usepackage{cite}
\usepackage{ulem}
\definecolor{dgreen}{rgb}{0, 0.8, 0.1}

\usepackage{breqn}

\begin{document}

\title{Joint Image De-noising and Enhancement for Satellite-Based SAR  \\
\author{\IEEEauthorblockN{Shahrokh Hamidi}\\
\IEEEauthorblockA{{Department of Electrical and Computer Engineering},
{University of Waterloo}\\
Waterloo, Ontario, Canada \\
shahrokh.hamidi@uwaterloo.ca}
}
}

\maketitle

\begin{tikzpicture}[remember picture, overlay]
      \node[font=\small] at ([yshift=-1cm]current page.north)  {This paper has been published in the 2024 IEEE International Conference on Aerospace Electronics and Remote Sensing Technology (ICARES). \copyright IEEE};
\end{tikzpicture}

\begin{abstract}
The reconstructed images from the Synthetic Aperture Radar (SAR) data suffer from multiplicative noise as well as low contrast level. 
These two factors impact the quality of the SAR images significantly and prevent any attempt to extract valuable information from the processed data. 
The necessity for mitigating these effects in the field of SAR imaging is of high importance. 
Therefore, in this paper, we address the aforementioned issues and propose a technique to handle these shortcomings simultaneously. In fact, we combine the de-noising and contrast enhancement processes into a unified algorithm.
The image enhancement is performed based on the Contrast Limited Adaptive Histogram Equalization (CLAHE) technique. 

The verification of the proposed algorithm is performed by experimental results based on the data that has been collected from the European Space Agency’s ERS-2 satellite which operates in strip-map mode. 
\end{abstract}

\section{Introduction}
With numerous applications for Synthetic Aperture Radar (SAR) imagery systems, the importance of producing images with high quality can not be underestimated \cite{Cumming, Curlander, Soumekh, Harger, Sullivan, Munson_Stripmap}. SAR imaging technique is considered to be one of the most important methods to produce high resolution images from the surface of the Earth. The all-weather and day-night capabilities as well as large area coverage of the Earth have made SAR imaging the subject for significant research in the scientific society \cite{Cumming, Curlander, Soumekh, Harger, Sullivan, Munson_Stripmap}.
However, despite all the attempts in the field of SAR image reconstruction, there are still shortcomings in addressing some of the crucial problems that exist with the reconstructed images. Due to the coherent nature of the SAR systems, the reconstructed images are contaminated with a type of multiplicative noise, known as speckle noise, which degrades the quality of the reconstructed images considerably \cite{Cumming}. Furthermore, the reconstructed images show low contrast, which reduces the quality of the images significantly. 

In this paper, we present the SAR image formation in detail. We then address the issue related to the speckle noise \cite{Cumming} and contrast enhancement \cite{img_proc_Gonzalez}. The common approach for speckle noise removal is multi-look processing in which the entire synthetic aperture is divided into several sub-apertures and then per each sub-aperture an image is created. The final image is the non-coherent average of the reconstructed images from the sub-apertures \cite{Cumming}. The main issue with the multi-look processing is the reduction of the resolution limit of the imagery system in the azimuth direction. Moreover, the multi-look processing approach can not be integrated to the image enhancement algorithm that we present in this paper. Therefore, we utilize median filtering method in order to alleviate the effect of the speckle noise \cite{img_proc_Gonzalez}. Median filtering is proved to be the appropriate approach to address the speckle noise issue since it alleviates the effect of the speckle noise significantly while the resolution limit in the azimuth direction remains intact.   
The common approach for the median filter implementation is time-consuming and computationally expensive. Instead, we utilize the histogram-based technique to implement the median filter \cite{Median_Fitelr_1} which is significantly faster and computationally more efficient compared to the common methods of implementing it. In addition, the histogram-based approach will pave the way for combining the median filtering and contrast enhancement process which is the main goal of this work. We utilize the Contrast Limited Adaptive Histogram Equalization (CLAHE) technique \cite{CLAHE1, CLAHE2} to enhance the contrast of the reconstructed image while the median filter is applied to the image to mitigate the effect of the speckle noise simultaneously. Combining these two methods and merging them into one unified technique will reduce the computational complexity of these algorithms compared to the case in which they are implemented separately. At the end,  we verify the effectiveness of the proposed algorithm by applying it to the reconstructed image from the raw data that has been collected from the European Space Agency’s ERS-2 satellite. The raw data has been published by NASA's Alaska Satellite Facility (ASF).

The organization of the paper is as follows. In Section \ref{Model Geometry}, we describe the geometry of the model and formulate the problem. In Section \ref{Wavenumber Algorithm}, we present the $\omega - k$ algorithm which is utilized for image reconstruction. In Section \ref{Image De-nosing and Enhancement}, we discuss the image de-noising as well as image enhancement processes. Section \ref{Experimental Results} has been dedicated to the experimental results, based on the data collected from the European Space Agency’s ERS-2 satellite, followed by the concluding remarks. 
\section{Model Geometry}\label{Model Geometry}
\begin{figure}
\centering
\begin{tikzpicture}[yshift=0.00001cm][font=\large]
\node(img1) {\includegraphics[height=6cm,width=8cm]{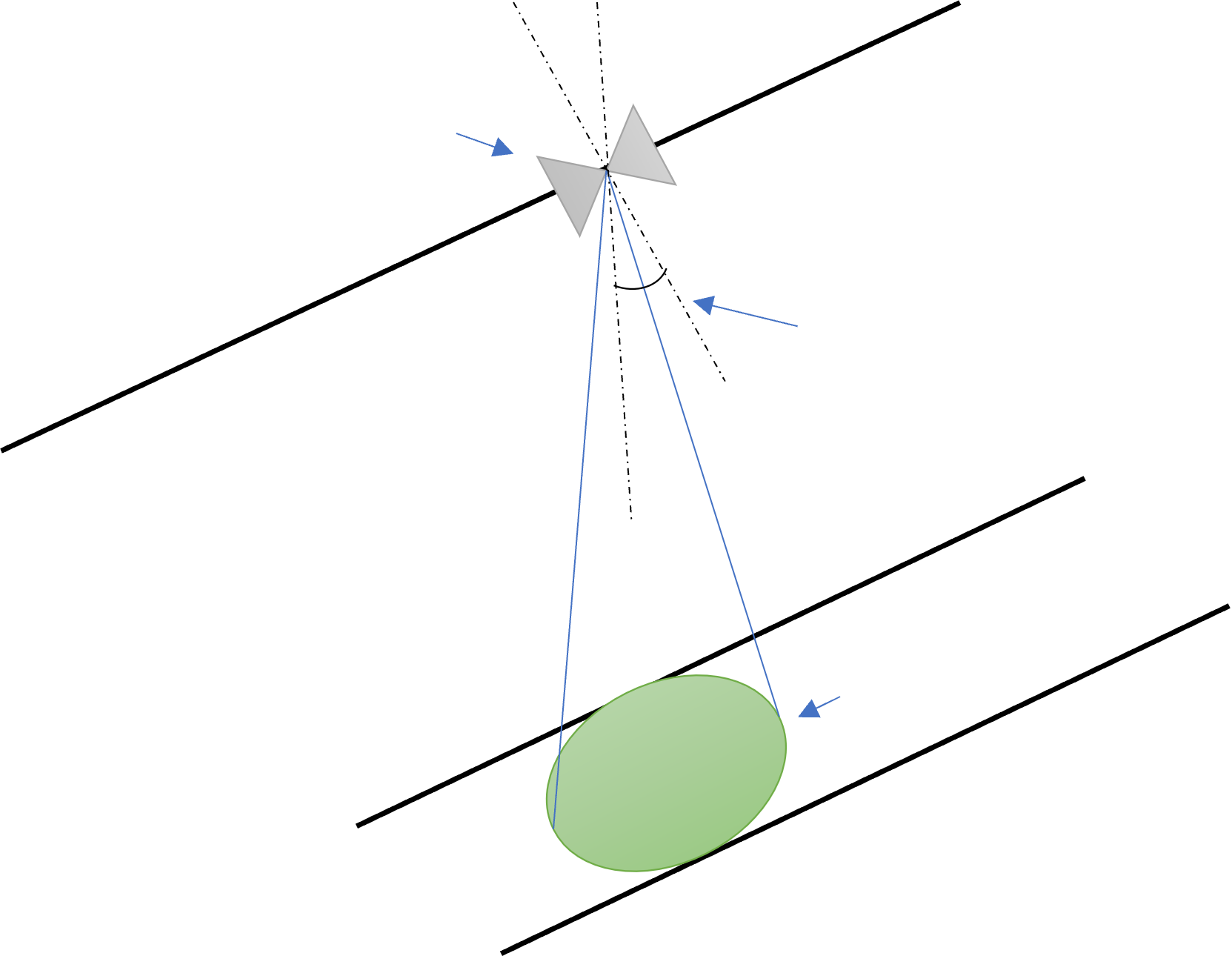}};
\node[left=of img1, node distance=0cm, rotate = 25, xshift=5cm, yshift=0.5cm,font=\color{black}] {{Satellite}};
\node[left=of img1, node distance=0cm, rotate = 25, xshift=5.7cm, yshift=-1.5cm,font=\color{black}]  {{$\theta_{\rm sq}$}};
\node[left=of img1, node distance=0cm, rotate = 25, xshift=8cm, yshift=-2.2cm,font=\color{black}]  {{Squint Angle}};
\node[left=of img1, node distance=0cm, rotate = 25, xshift=8.8cm, yshift=-4cm,font=\color{black}]  {{Antenna Footprint}};
\end{tikzpicture}
\caption{{The geometry of the model which shows a satellite-based SAR system in strip-map mode with non-zero squint angle. }
\label{fig:Model_Geometry}}
\end{figure}
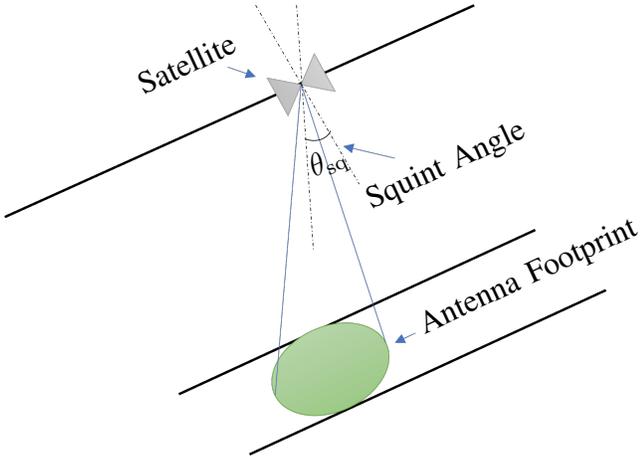

The geometry of the model has been depicted in Fig.~\ref{fig:Model_Geometry}.
The signal received at the location of the receiver after being reflected  from a point reflector, is a chirp signal which is modeled as \cite{Cumming}
\begin{dmath}
\label{RX_signal_0}
 s(t, \eta) = \sigma \; w_r\left(t - \frac{2R(\eta)}{c}\right)w_a(\eta - \eta_c) \times
 e^{\displaystyle  j 2 \pi f_c \left(t - \frac{2R(\eta)}{c}\right) + j \pi \beta {\left(t - \frac{2R(\eta)}{c}\right)}^2},
\end{dmath}
where $f_c$ is the carrier frequency and the parameter $\beta$ is given as $b/T$, in which $b$ and $T$ stand for the bandwidth and the chirp time, respectively. In addition, $w_r$ is a rectangular window with length $T$ and $t$ is referred to as fast-time parameter.
Furthermore, the parameter $\sigma$ is the complex radar cross section for the point reflector, $\eta$ is referred to as slow-time parameter and $R(\eta)$ is the instantaneous radial distance between the radar and the target and is described as  $R(\eta) = \sqrt{R^2_0 + v^2(\eta - \eta_c)^2}$. Moreover, $w_a$ is a rectangular window with its width equal to the synthetic aperture length divided by $v$ which is the orbital speed of the satellite.
Additionally, the parameter $\eta_c$ is the slow-time moment at which the peak of the main-lobe of the antenna illuminates the point reflector. The Doppler frequency at $\eta_c$ is referred to as Doppler centroid frequency and is written as $f_{\rm dc}$ which plays a crucial role in SAR image reconstruction \cite{Johnson_fdc, Madsen, Bamler_fdc, Shu_fdc_MLBF, Shu_fdc_MLBF_2}.    
\section{$\omega-k$ Algorithm}\label{Wavenumber Algorithm}
In this section, we present the $\omega-k$ algorithm \cite{Cumming_wk, Bamler_RD_wk, Cumming}, which is also referred to as Wavenumber method as well as Range Migration Algorithm (RMA).

In the $\omega-k$ algorithm, the range compression, the range cell migration (RCM) compensation, and the azimuth localization are all performed in the 2D frequency domain.
The first step in implementing the  $\omega-k$ algorithm is to take a 2D Fourier transform from (\ref{RX_signal_0}) which upon using the Method Of Stationary Phase (MOSP) \cite{Optics} we obtain 
\begin{dmath}
\label{2D_FFT}
 S(f_t, f_\eta) = \sigma \; W_r(f_t)W_a(f_{\eta} - f_{\eta_c}) \times 
 e^{\displaystyle  -j 4 \pi \frac{R_0}{c}\sqrt{(f_c+f_t)^2 - \frac{c^2f^2_{\eta}}{4v^2}} - \frac{\pi f^2_t}{\beta}},
\end{dmath}
where $f_t$ and $f_{\eta}$ are the fast-time and slow-time frequency components, respectively.
The last term in the phase of (\ref{2D_FFT}) represents a chirp signal in the fast-time frequency domain and is compensated by multiplying the signal by $e^{\displaystyle \frac{\pi f^2_t}{\beta}}$.  Basically, by compensating for this term, we are performing de-chirping process which results in energy localization for the point target in the range direction.  
We then multiply (\ref{2D_FFT}) by the reference phase function defined as
\begin{eqnarray}
\label{Ref_Phase}
 S_{\rm ref}(f_t, f_\eta) = e^{\displaystyle  j 4 \pi \frac{R_{\rm ref}}{c}\sqrt{(f_c+f_t)^2 - \frac{c^2f^2_{\eta}}{4v^2}}}.
\end{eqnarray}
The physical meaning of eliminating the phase term given in (\ref{Ref_Phase}) is that, by performing this task, we propagate the energy of the point target back to the plane that has been located at $R_{\rm ref}$. Upon implementing this task, we obtain
\begin{dmath}
\label{2D_FFT_c}
 S_c(f_t, f_\eta) = \sigma \; W_r(f_t)W_a(f_{\eta} - f_{\eta_c}) \times  
 e^{\displaystyle  -j 4 \pi \frac{R_0-R_{\rm ref}}{c}\sqrt{(f_c+f_t)^2 - \frac{c^2f^2_{\eta}}{4v^2}}}.
\end{dmath}
For a target located at $R_{\rm ref}$ the phase component in (\ref{2D_FFT_c}) is removed completely and upon taking an inverse 2D Fourier transform, we will localize the energy of the target in both the range and azimuth directions.
For targets located at ranges other than $R_{\rm ref}$, however, there is a residual phase term in (\ref{2D_FFT_c}).

The next step in the  $\omega-k$ algorithm, is to linearize the phase term given in (\ref{2D_FFT_c}) which is performed using Stolt interpolation \cite{Bamler_RD_wk, Cumming_wk, Cumming}. The Stolt interpolation procedure is given as $\sqrt{(f_c+f_t)^2 - \frac{c^2f^2_{\eta}}{4v^2}}\rightarrow f_c + f^{\prime}_t$. After performing Stolt interpolation, we obtain
\begin{dmath}
\label{2D_FFT_cs}
 S_{cs}(f^{\prime}_t, f_\eta) =  \sigma \; W_r(f^{\prime}_t)W_a(f_{\eta} - f_{\eta_c}) \times 
 e^{\displaystyle  -j 4 \pi \left( \frac{R_0-R_{\rm ref}}{c} \right) (f_c+f^{\prime}_t)}.
\end{dmath}
Finally, by taking a 2D inverse Fourier transform from (\ref{2D_FFT_cs}) using MOSP, we localize the energy of the target in both the range and azimuth directions.
To remove the effect of the $R_{\rm ref}$, prior to taking the Fourier transform, we multiply (\ref{2D_FFT_cs}) by $e^{ \displaystyle  -j 4 \pi \frac{R_{\rm ref}}{c}f^{\prime}_t}$ and then perform the inverse 2D Fourier transform to achieve the reconstructed image as
\begin{dmath}
\label{O_K_Image}
s_c(t,\eta) =  \sigma \; p_r(t - 2\frac{R_0}{c})p_a(\eta)\times 
 e^{\displaystyle  j 2 \pi f_{\eta_c}\eta}\;e^{\displaystyle  -j 4 \pi \left( \frac{R_0-R_{\rm ref}}{c} \right) f_c t}.
\end{dmath}
In the $\omega-k$ algorithm, the RCM is compensated in the 2D frequency domain. Therefore, the dependency of the RCM on range is ignored. Performing the RCM in the 2D frequency domain instead of the range Doppler domain is a compromise between speed and accuracy \cite{Cumming_wk, Bamler_RD_wk, Cumming}.
\section{Image De-nosing and Enhancement}\label{Image De-nosing and Enhancement}
In this section, we address  the image de-noising process. We attempt to mitigate the effect of the speckle noise which is due to the coherent nature of the SAR systems. By being aware of the fact that, the speckle noise is a multiplicative noise in nature, we apply the median filter technique in order to eliminate its effect on the reconstructed image \cite{img_proc_Gonzalez}. The behavior of the speckle noise is similar to the impulsive noise. Therefore, the samples of the speckle noise can be considered as outliers and that is the reason that median filter can reduce its effect significantly. However, the speckle noise is not the only issue with the reconstructed images. The contrast of the SAR images is low which can hinder our ability to extract necessary information from the images.    

It should be noted that, the common method for speckle noise removal in SAR imaging, is multi-look processing in which the aperture is divided into several sub-apertures and per each one of them an image is created \cite{Cumming}. The final image is created by performing non-coherent averaging over all the images that have been produced from the sub-apertures. The main issue with the multi-look processing is that it decreases the resolution in the azimuth direction. Applying median filter, however, can mitigate the adverse effect of the speckle noise while the resolution in the azimuth direction remains intact. 
Another problem with the multi-look processing is its incompatibility with the image enhancement approach that we utilize in this paper. 

One of the techniques that is offered for increasing the contrast of images is histogram equalization \cite{img_proc_Gonzalez}. In essence, the histogram equalization attempts to spread the intensity values. However, this can lead to images with pixels that are excessively bright or significantly dark. To tackle this issue, we will implement the CLAHE algorithm \cite{CLAHE1, CLAHE2} which is an advanced version of the adaptive histogram equalization. In CLAHE method, the image is divided into smaller blocks and the histogram equalization is applied to each block independently and the histogram clipping is included in the algorithm.  

In this paper, we will combine the CLAHE algorithm with the median filter and create a unified package to de-noise the image and enhance its contrast simultaneously. Compared to the case in which the median filter and the CLAHE are implemented separately, the combined approach reduces the computational complexity.
Moreover, it should be emphasized that, the technique that we utilize for image enhancement is not capable of handling the speckle noise. The effect of speckle noise is mitigated by utilizing the median filter. 

Instead of implementing the median filter by sorting the data and finding the middle value which is a time-consuming process and demands high computational complexity, we utilize a histogram-based approach which is significantly faster and more efficient \cite{Median_Fitelr_1}. Furthermore, the histogram-based approach will pave the way to combine the median filter implementation with the CLAHE algorithm. 
\section{Experimental Results}\label{Experimental Results}
\begin{figure}
\centering
\begin{tikzpicture}[yshift=0.00001cm][font=\large]
\node(img1) {\includegraphics[height=6cm,width=8cm]{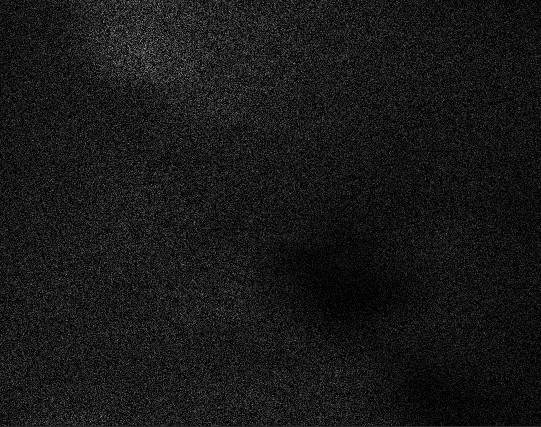}};
\node[left=of img1, node distance=0cm, rotate = 90, xshift=1.3cm, yshift=-0.7cm,font=\color{black}] {{Slant-Range}};
\node[below=of img1, node distance=0cm, xshift=0cm, yshift=1.1cm,font=\color{black}] {{Along-Track}};
\end{tikzpicture}
\caption{{The image presents the absolute value of the raw data which has been collected from the satellite. The data is in In-phase and Quadrature format.}
\label{fig:img_raw}}
\end{figure}

\begin{figure}
\centering
\begin{tikzpicture}[yshift=0.00001cm][font=\large]
\node(img1) {\includegraphics[height=4cm,width=6cm]{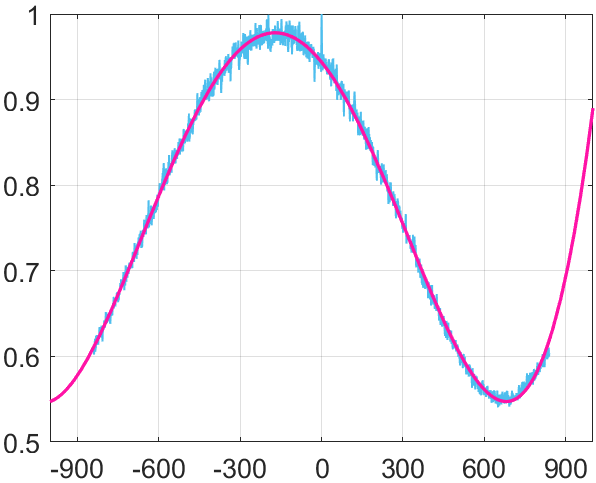}};
\node[left=of img1, node distance=0cm, rotate = 90, xshift=2cm, yshift=-0.85cm,font=\color{black}] {{Normalized Power}};
\node[below=of img1, node distance=0cm, xshift=0.4cm, yshift=1.1cm,font=\color{black}] {{Frequency [Hz]}};
\end{tikzpicture}
\caption{{The result for the amplitude-based Doppler centroid frequency estimation. The estimated value is $\rm f_{dc} = -169 \; Hz$. The estimation is based on averaging over all 4912 range cells.}
\label{fig:Doppler_centroid}}
\end{figure}

\begin{figure}
\centering
\begin{tikzpicture}[yshift=0.00001cm][font=\small]
\node(img1) {\includegraphics[height=6cm,width=6cm]{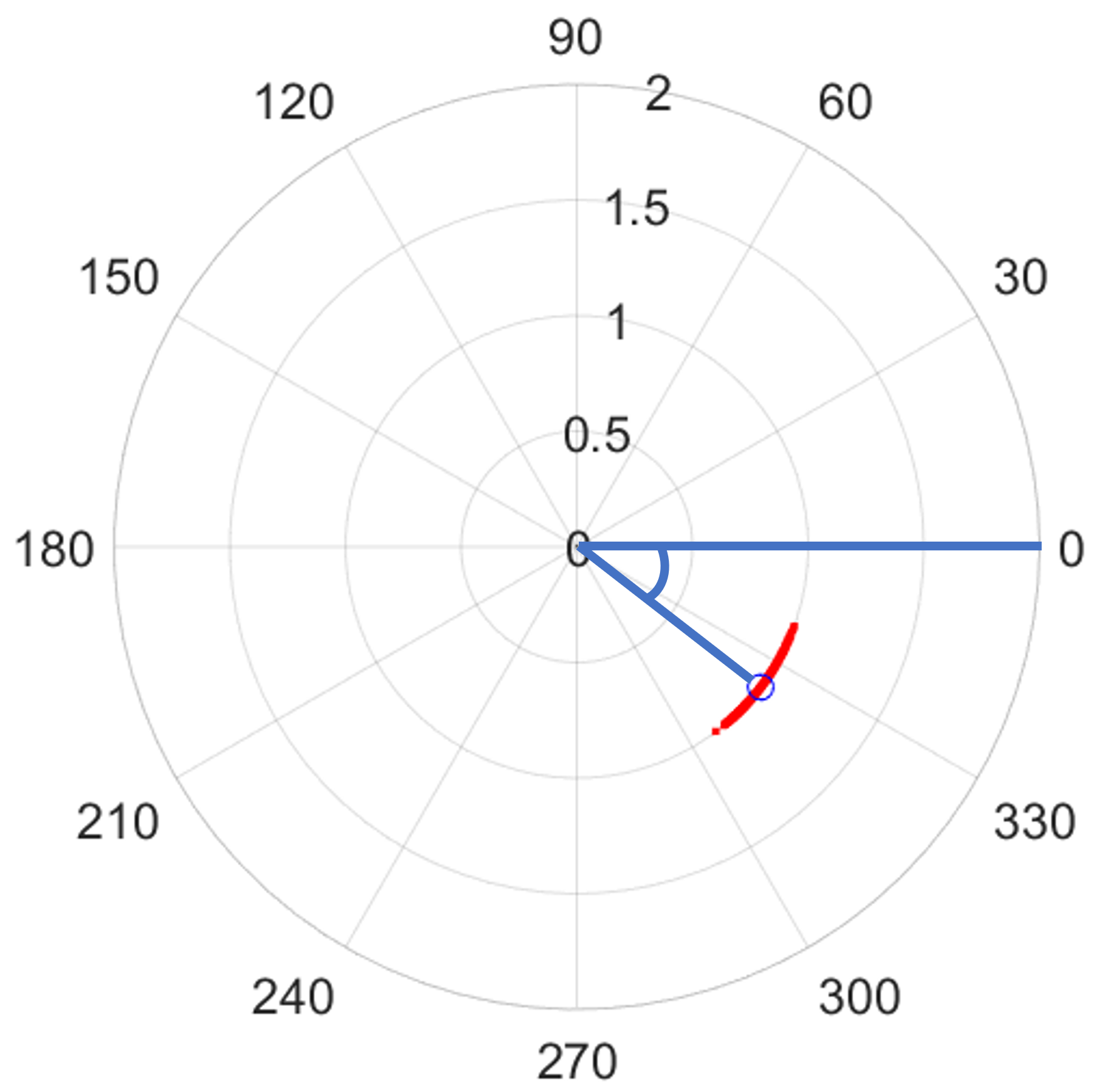}};
\node[left=of img1, node distance=0cm, rotate = 0, xshift=6.3cm, yshift=-0.2cm,font=\color{black}]  {{$\rm -0.66 \; rad$}};
\end{tikzpicture}
\caption{{The result for the phase-based Doppler centroid frequency estimation using ACCC method. The estimated value is $\rm f_{dc} = -176 \; Hz$. The estimation is based on averaging over all 4912 range cells.}
\label{fig:Doppler_centroid_phase}}
\end{figure}

\begin{figure}
\centering
\begin{tikzpicture}[yshift=0.00001cm][font=\large]
\node(img1) {\includegraphics[height=6cm,width=8cm]{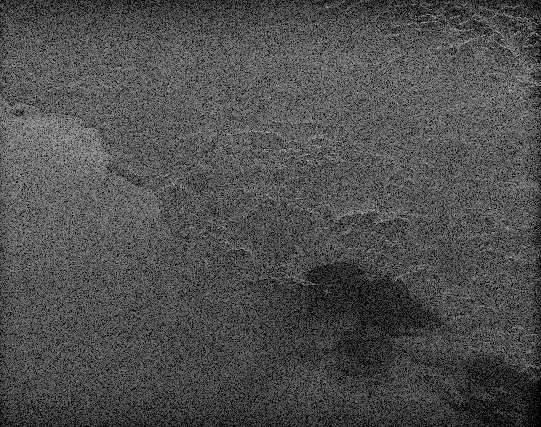}};
\node[left=of img1, node distance=0cm, rotate = 90, xshift=1.3cm, yshift=-0.7cm,font=\color{black}] {{Slant-Range}};
\node[below=of img1, node distance=0cm, xshift=0cm, yshift=1.1cm,font=\color{black}] {{Along-Track}};
\end{tikzpicture}
\caption{{The reconstructed image from the raw data, presented in Fig.~\ref{fig:img_raw}, based on the $\omega-k$ algorithm.}
\label{fig:img_noisy}}
\end{figure}

\begin{figure}
\centering
\begin{tikzpicture}[yshift=0.00001cm][font=\small]
  \node (img1)  {\includegraphics[height=6cm,width=8cm]{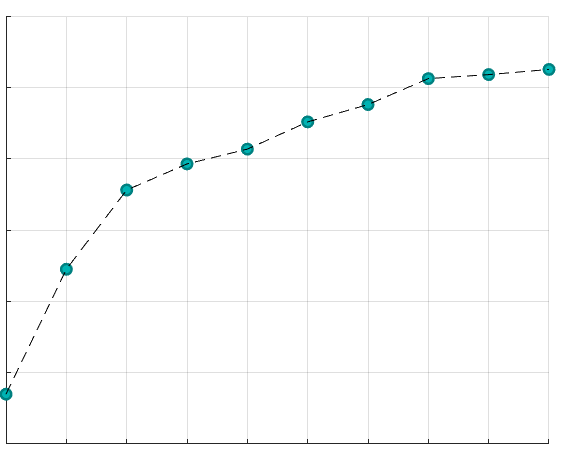}};
  \node[left=of img1, node distance=0cm, rotate = 90, xshift=1.2cm, yshift=-0.5cm,font=\color{black}] {{\large  PSNR [dB]}};
 \node[left=of img1, node distance=0cm, rotate = 90,  xshift=-2.95cm, yshift=-1.25cm,font=\color{black}] {{$\rm p=q=2$ }};
  \node[left=of img1, node distance=0cm, rotate = 90,  xshift=-2.95cm, yshift=-2.09cm,font=\color{black}] {{$\rm p=q=4$}};
  \node[left=of img1, node distance=0cm, rotate = 90,  xshift=-2.95cm, yshift=-2.92cm,font=\color{black}] {{$\rm p=q=6$}};
  \node[left=of img1, node distance=0cm, rotate = 90,  xshift=-2.95cm, yshift=-3.8cm,font=\color{black}] {{$\rm p=q=8$ }};
  \node[left=of img1, node distance=0cm, rotate = 90,  xshift=-2.95cm, yshift=-4.65cm,font=\color{black}] {{$\rm p=q=10$}};
  \node[left=of img1, node distance=0cm, rotate = 90,  xshift=-2.95cm, yshift=-5.5cm,font=\color{black}] {{$\rm p=q=12$}};
  \node[left=of img1, node distance=0cm, rotate = 90,  xshift=-2.95cm, yshift=-6.35cm,font=\color{black}] {{$\rm p=q=14$}};
  \node[left=of img1, node distance=0cm, rotate = 90,  xshift=-2.95cm, yshift=-7.2cm,font=\color{black}] {{$\rm p=q=16$}};
  \node[left=of img1, node distance=0cm, rotate = 90,  xshift=-2.95cm, yshift=-8.05cm,font=\color{black}] {{$\rm p=q=18$}};
\node[left=of img1, node distance=0cm, rotate = 90,  xshift=-2.95cm, yshift=-8.85cm,font=\color{black}] {{$\rm p=q=20$}};

\node[left=of img1, node distance=0cm, xshift=1.2cm, yshift=-2.8cm,font=\color{black}] {{10}};
\node[left=of img1, node distance=0cm, xshift=1.2cm, yshift=-1.9cm,font=\color{black}] {{15}};
\node[left=of img1, node distance=0cm, xshift=1.2cm, yshift=-1cm,font=\color{black}] {{20}};
 \node[left=of img1, node distance=0cm, xshift=1.2cm, yshift=-0.02cm,font=\color{black}] {{25}};
\node[left=of img1, node distance=0cm, xshift=1.2cm, yshift=0.9cm,font=\color{black}] {{30}};
\node[left=of img1, node distance=0cm, xshift=1.2cm, yshift=1.84cm,font=\color{black}] {{35}};
\node[left=of img1, node distance=0cm, xshift=1.2cm, yshift=2.75cm,font=\color{black}] {{40}};
\end{tikzpicture}
\caption{The PSNR versus different values of the parameters of the median filter, namely  $\rm p$ and $\rm q$.}
\label{fig:PSNR}
\end{figure}

The experimental results that we present in this section, is based on the data collected from the European Space Agency’s ERS-2 satellite which operates in strip-map mode. The ERS-2 data-set has been published by the NASA's Alaska Satellite Facility (ASF).  
The specifications for the satellite have been given in Table.\ref{Tab:Table}.
\begin{table}
  \centering
  \caption{ERS-2's Parameters}\label{Tab:Table}
  \begin{center}
    \begin{tabular}{| l | l | l |}

    \hline
    Parameters &  & Values  \\ \hline
    Center frequency ($\rm GHz$) & $\rm f_c$ & $5.3$ \\ \hline
    Radar sampling rate ($\rm MHz$) & $\rm Fr$ & $18.97$  \\ \hline
    Pulse repetition frequency ($\rm Hz$) & $\rm PRF$ & $1679$ \\ \hline
    Slant range of first radar sample ($\rm km$) & $R_0$ & $830.9$  \\ \hline
    FM rate of the pulsed radar (${\rm MHz}/\mu s$) & $ \beta$     &  $4.19$   \\ \hline
    Chirp duration ($\mu s$)  &  $\rm T$  &  $37.1$   \\ \hline
    Satellite velocity ($\rm m/s$) &   $\rm v$  &  $7120$    \\ \hline
    Bandwidth ($\rm MHz$)  &  $\rm b$   & $15.545$ \\ \hline
    \hline
    \end{tabular}
\end{center}
\end{table}
Fig.~\ref{fig:img_raw} illustrates the absolute value of the raw data captured by the satellite. The data is in In-phase (I) and Quadrature (Q) format. The number of bins in the range and along-track directions are $4912$ and $27975$, respectively. 
Fig.~\ref{fig:Doppler_centroid} presents the result for the amplitude-based Doppler centroid frequency estimation process \cite{Cumming, Johnson_fdc, Madsen, Bamler_fdc, Shu_fdc_MLBF, Shu_fdc_MLBF_2}. In order to reduce the effect of the noise, we have performed coherent averaging over all 4912 range cells which is the total number of range cells available. It should be noted that, the averaging process ignores the dependency of the Doppler centroid on range. However, as the reconstructed image shows, the dependency is not highly significant. To estimate the maximum of the graph, shown in Fig.~\ref{fig:Doppler_centroid}, we have performed polynomial fitting which, as a result, the estimated value for the Doppler centroid frequency is obtained as $\rm f_{dc} = -169 \; Hz$. Subsequently, the squint angle is calculated as $\rm -0.038^o$ which implies that the satellite is operating under low squint angle.  
To obtain a robust estimation for the Doppler centroid frequency, in addition to the amplitude-based approach, we have implemented a phase-based approach as well. Fig.~\ref{fig:Doppler_centroid_phase} presents the result for the phase-based Doppler centroid frequency estimation based on the Average Cross Correlation Coefficient (ACCC) method which is described as \cite{Cumming} 
\begin{eqnarray}
\label{fdc_phased}
\bar C  =  \sum_{t} \sum_{\eta} s^*_{\rm rc}(t, \eta)s_{\rm rc}(t, \eta + \delta \eta),
\end{eqnarray}
in which  $s_{\rm rc}(t, \eta)$ represents the range compressed signal. Subsequently, the Doppler centroid frequency is estimated as ${\rm  f_{ dc}} = \frac{\rm PRF}{2\pi} \angle \bar{C}$. The angle is calculated as $\rm \angle \bar{C} = -0.66 \; rad$ which results in the estimated value for the Doppler centroid frequency as $\rm f_{dc} = -176 \; Hz$. 
Finally, we take the average value of the results for the amplitude-based method and the phase-based technique and obtain $\rm -172.5 \; Hz$ as an estimate for the Doppler centroid frequency.   
The reconstructed image based on the $\omega-k$ algorithm has been shown in Fig.~\ref{fig:img_noisy}. As it is evident from Fig.~\ref{fig:img_noisy}, the speckle noise has covered all the details of the image.
To tackle the issue of the speckle noise, we apply a median filter to the reconstructed image which has been presented in Fig.~\ref{fig:img_noisy}. The primary factor for the success of the median filter is the fact that the noisy samples behave as outliers which can be removed by median filter efficiently.  
In order to estimate the optimum values for the parameters of the median filter, namely $\rm p$ and $\rm q$ which represent the width and height of the 2D filter, we have presented the Peak Signal to Noise Ratio (PSNR) of the de-noised image versus $\rm p$ and $\rm q$ in Fig.~\ref{fig:PSNR}. The PSNR is defined as $ \rm PSNR = \displaystyle 10 \log \frac{\rm |img|^2_{max}}{\rm MSE}$ in which  the Mean Squared Error (MSE) is described as $\rm MSE = \displaystyle \rm \frac{\sum_{i=1,j=1}^{M, N} {\left(\rm |img_{noisy}[i,j]| - |img[i,j]|\right)}^2}{MN}$. Additionally,  $\rm |img|_{max} = 255$ and $\rm |img| = \sqrt{I^2 + Q^2}$. Furthermore, $\rm |img_{noisy}|$ represents the noisy image. From  Fig.~\ref{fig:PSNR}, it is evident that by increasing the size of the median filter, the PSNR enhances until it reaches a plateau at $\rm 35.61 \;dB$ for $\rm p=q=16$. This demonstrates that the optimum values for the parameters of the median filter are $\rm p=q=16$. Increasing the value of the  parameters beyond this limit will result in a slight improvement in the PSNR. However, it increases the computational complexity of the algorithm. Moreover, image blurring occurs as well. Therefore,  the value $16$ is selected for both $\rm p$ and $\rm q$.   

\begin{figure}
\centering
\begin{tikzpicture}[yshift=0.00001cm][font=\large]
\node(img1) {\includegraphics[height=6cm,width=8cm]{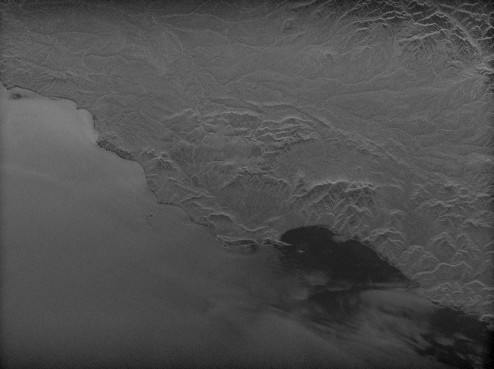}};
\node[left=of img1, node distance=0cm, rotate = 90, xshift=1.3cm, yshift=-0.7cm,font=\color{black}] {{Slant-Range}};
\node[below=of img1, node distance=0cm, xshift=0cm, yshift=1.1cm,font=\color{black}] {{Along-Track}};
\end{tikzpicture}
\caption{{The reconstructed image followed by speckle noise removal.}
\label{fig:img_de-noised}}
\end{figure}

\begin{figure}
\centering
\begin{tikzpicture}[yshift=0.00001cm][font=\large]
\node(img1) {\includegraphics[height=6cm,width=8cm]{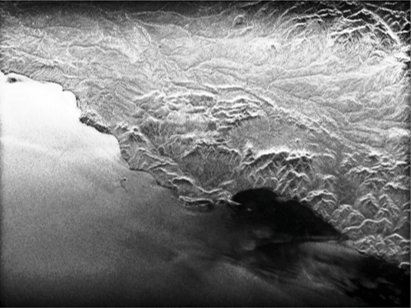}};
\node[left=of img1, node distance=0cm, rotate = 90, xshift=1.3cm, yshift=-0.7cm,font=\color{black}] {{Slant-Range}};
\node[below=of img1, node distance=0cm, xshift=0cm, yshift=1.1cm,font=\color{black}] {{Along-Track}};
\end{tikzpicture}
\caption{{The result of applying the histogram equalization to the image shown in Fig.~\ref{fig:img_de-noised}.}
\label{fig:hist_eq}}
\end{figure}

\begin{figure}
\centering
\begin{tikzpicture}[yshift=0.00001cm][font=\large]
  \node (img1)  {\includegraphics[height=5.5cm,width=8cm]{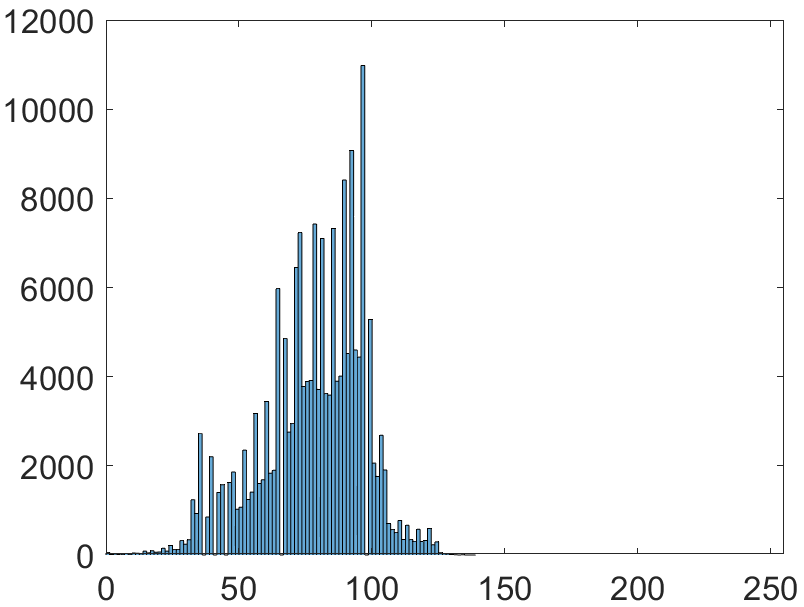}};
\node[left=of img1, node distance=0cm, rotate = 90, xshift=1.3cm, yshift=-0.9cm,font=\color{black}] {{Frequency}};
\node[below=of img1, node distance=0cm, xshift=0.5cm, yshift=1.15cm,font=\color{black}] {{Gray Level}}; 
\node[below=of img1, node distance=0cm, xshift=0.5cm, yshift=0.7cm,font=\color{black}] {{(a)}};
\end{tikzpicture}
\begin{tikzpicture}[yshift=0.00001cm][font=\large]
\node (img1)  {\includegraphics[height=5.5cm,width=8cm]{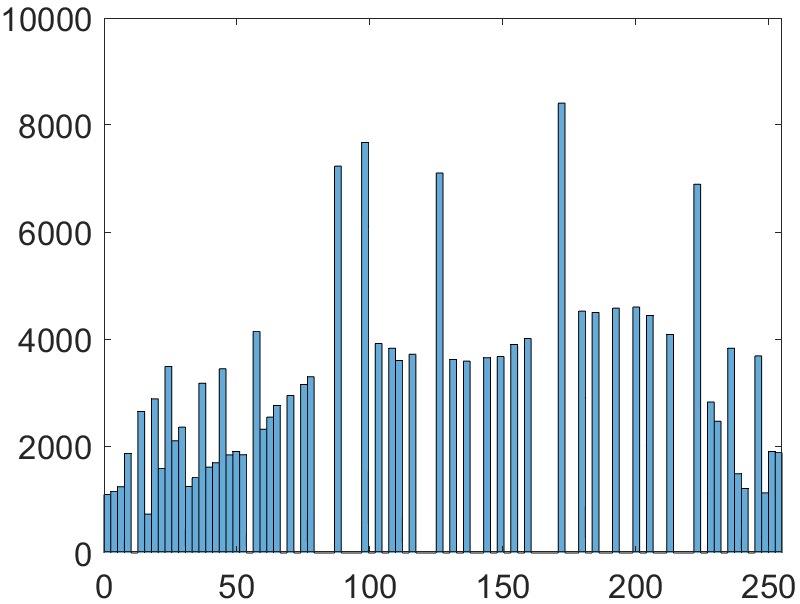}};
\node[left=of img1, node distance=0cm, rotate = 90, xshift=1.3cm, yshift=-0.9cm,font=\color{black}] {{Frequency}};
\node[below=of img1, node distance=0cm, xshift=0.5cm, yshift=1.15cm,font=\color{black}] {{Gray Level}}; 
\node[below=of img1, node distance=0cm, xshift=0.5cm, yshift=0.7cm,font=\color{black}] {{(b)}};
\end{tikzpicture}
\caption{{The histogram for the images presented in, a) Fig.~\ref{fig:img_de-noised}, b) Fig.~\ref{fig:hist_eq}.}
\label{fig:hist_eq_graph}}
\end{figure}

Fig.~\ref{fig:img_de-noised} illustrates the de-noised image. From Fig.~\ref{fig:img_de-noised}, we can clearly see the fine details of the image with clarity. The edges have been preserved and image smearing has not been occurred. 
In spite of the fact that the de-noised image is of high quality and the details and fine structures can easily be seen, the contrast of the image is low. To resolve this problem and enhance the contrast of the image, we apply the histogram equalization technique.  
Fig.~\ref{fig:hist_eq} depicts the result of applying the histogram equalization to the image given in Fig.~\ref{fig:img_de-noised}. 
In order to observe the effect of applying the histogram equalization method to the image, the histogram of the image has been shown in Fig.~\ref{fig:hist_eq_graph}.
In Fig.~\ref{fig:hist_eq_graph}-(a),  the histogram for the image presented in Fig.~\ref{fig:img_de-noised} has been depicted. Moreover, Fig.~\ref{fig:hist_eq_graph}-(b) illustrates the histogram of the image following the utilization of the histogram equalization technique. 
The concentration of the values of the pixels at a specific part of the graph presented in Fig.~\ref{fig:hist_eq_graph}-(a) implies that the image is of low contrast. In Fig.~\ref{fig:hist_eq_graph}-(b), however, the result demonstrates the effect of applying the histogram equalization method to the image. 

\begin{figure}
\centering
\begin{tikzpicture}[yshift=0.00001cm][font=\large]
\node(img1) {\includegraphics[height=6cm,width=8cm]{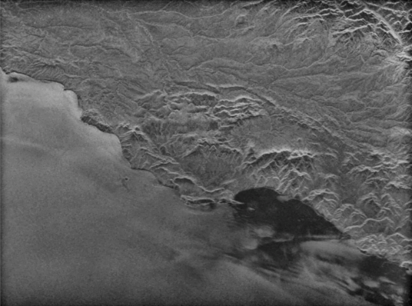}};
\node[left=of img1, node distance=0cm, rotate = 90, xshift=1.3cm, yshift=-0.7cm,font=\color{black}] {{Slant-Range}};
\node[below=of img1, node distance=0cm, xshift=0cm, yshift=1.1cm,font=\color{black}] {{Along-Track}};
\end{tikzpicture}
\caption{{The result of applying the joint histogram-based median filtering and the CLAHE method to the reconstructed image presented in Fig.~\ref{fig:img_noisy}.}
\label{fig:hist_eq_adv}}
\end{figure}

\begin{figure}
\centering
\begin{tikzpicture}[yshift=0.00001cm][font=\large]
  \node (img1)  {\includegraphics[height=5.5cm,width=8cm]{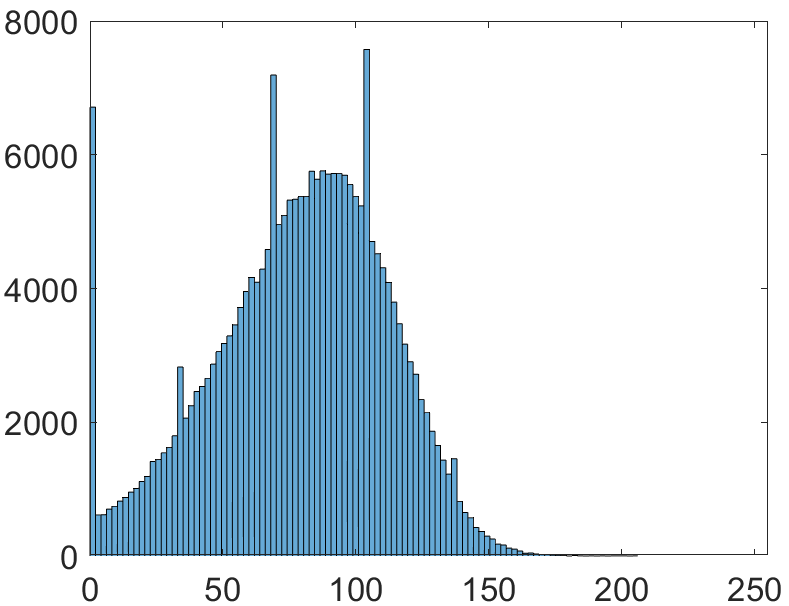}};
\node[left=of img1, node distance=0cm, rotate = 90, xshift=1.3cm, yshift=-0.9cm,font=\color{black}] {{Frequency}};
\node[below=of img1, node distance=0cm, xshift=0.5cm, yshift=1.15cm,font=\color{black}] {{Gray Level}}; 
\node[below=of img1, node distance=0cm, xshift=0.5cm, yshift=0.7cm,font=\color{black}] {{(a)}};
\end{tikzpicture}
\begin{tikzpicture}[yshift=0.00001cm][font=\large]
\node (img1)  {\includegraphics[height=5.5cm,width=8cm]{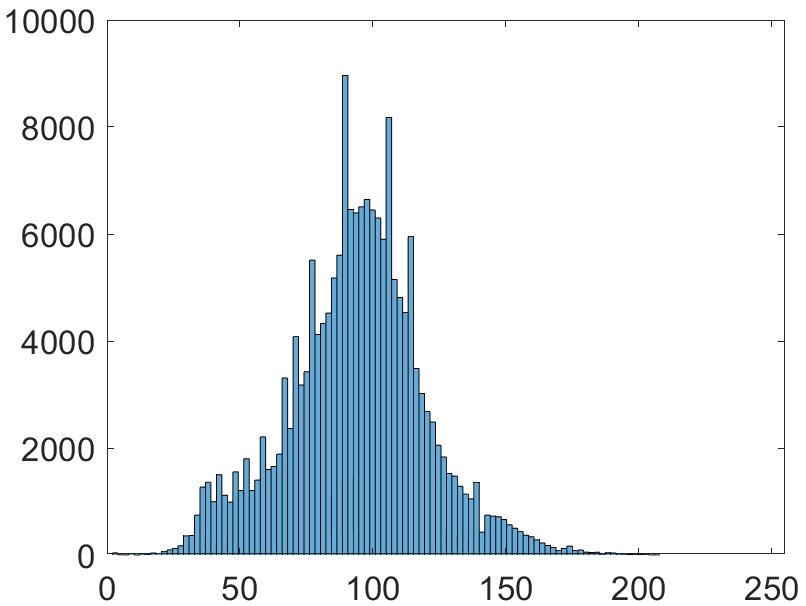}};
\node[left=of img1, node distance=0cm, rotate = 90, xshift=1.3cm, yshift=-0.9cm,font=\color{black}] {{Frequency}};
\node[below=of img1, node distance=0cm, xshift=0.5cm, yshift=1.15cm,font=\color{black}] {{Gray Level}}; 
\node[below=of img1, node distance=0cm, xshift=0.5cm, yshift=0.7cm,font=\color{black}] {{(b)}};
\end{tikzpicture}

\caption{{The histogram for the images presented in, a) Fig.~\ref{fig:img_noisy}, b) Fig.~\ref{fig:hist_eq_adv}.}
\label{fig:hist_eq_graph_adv}}
\end{figure}

It is worth noticing that despite of achieving higher quality  for the image after applying the histogram equalization, some of the pixels of the image are either excessively bright or extremely dark.  A possible way to solve this issue is to implement the CLAHE method. In fact, we can divide the image into smaller blocks and apply the histogram equalization to these blocks. Furthermore, per each block we perform the histogram-based median filtering and, as a result, we perform the speckle noise removal process as well as the contrast enhancement simultaneously.   The size of the blocks is $16 \times 16$ which is based on the analysis presented in Fig.~\ref{fig:PSNR}. 
In Fig.~\ref{fig:hist_eq_adv}, the result of applying the joint histogram-based median filtering and the CLAHE method to the reconstructed image, presented in Fig.~\ref{fig:img_noisy}, has been illustrated.
As it is evident from Fig.~\ref{fig:hist_eq_adv}, the effect of the speckle noise has been alleviated and the image shows high quality contrast. In addition, in contrast to the result which has been shown in Fig.~\ref{fig:hist_eq}, the image illustrated in Fig.~\ref{fig:hist_eq_adv}, does not suffer from pixels with excessive brightness or extreme low intensity. 
Moreover, Fig.~\ref{fig:hist_eq_graph_adv} shows the histogram for the image presented in Fig.~\ref{fig:hist_eq_adv}. From comparing the two histograms given in Fig.~\ref{fig:hist_eq_graph_adv}, we can observe the spread of the values which results in an image with higher contrast. 
%
%
%
\section{conclusion}\label{conclusion}
In this paper, we presented the SAR image reconstruction process in detail and then focused on speckle noise removal and contrast enhancement process. Next, we combined the image de-noising technique and the contrast enhancement method into a unified approach. Finally, the effectiveness of the proposed approach was verified by utilizing the experimental data and the results were presented.  

We firmly believe that, the joint image de-noising and adaptive contrast enhancement approach presented in this paper, can improve the quality of the SAR images considerably and reduce the computational complexities of the process significantly.

\begin{IEEEbiography}[{\includegraphics[width=1in,height=1.25in,clip,keepaspectratio]{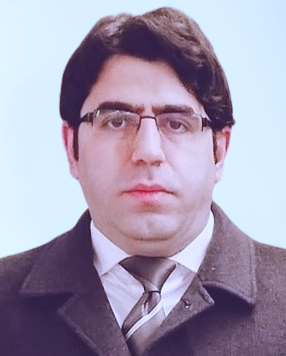}}]{Shahrokh Hamidi} was born in 1983, in Iran. He received his B.Sc., M.Sc., and Ph.D. degrees all in Electrical and Computer Engineering. He is with the faculty of Electrical and Computer Engineering at the University of Waterloo, Waterloo, Ontario, Canada. His current research areas include statistical signal processing, mmWave imaging, Terahertz imaging, image processing, system design,  multi-target tracking, wireless communication, machine learning, optimization, and array processing.
\end{IEEEbiography}


\end{document}